\documentclass[%
reprint, %if you want to see it in big to better proofreading, select preprint and not reprint
superscriptaddress,
nofootinbib,
 amsmath,amssymb,
 aps,
floatfix,
]{revtex4-2}

\usepackage{comment}
\usepackage{amsmath,amssymb}%,graphicx}
\usepackage[dvips, pdftex]{graphicx}
\usepackage[tight,footnotesize]{subfigure}
\usepackage{float}
\usepackage{wrapfig}
\usepackage{mathtools}
\usepackage[utf8]{inputenc}
\usepackage{graphicx}
\usepackage{physics}
\graphicspath{ {./images/} }
\usepackage{xcolor} % Allows defining custom colors
\definecolor{bblue}{RGB}{24, 116, 148}
\usepackage[colorlinks=true, linkcolor=bblue, citecolor=bblue, urlcolor=bblue]{hyperref}

\usepackage[many]{tcolorbox}  
\newtcolorbox{boxE}{
  enhanced,
  colback=white,
  colframe=black,
  boxrule=0.6pt,               
  sharp corners, width=0.31\textwidth,
  height=0.18\textwidth, 
 boxrule = 0pt, % clearing the default rule
    borderline = {0.75pt}{0pt}, % outer line
    borderline = {0.75pt}{0.06cm}, % inner line,
}

\newtcolorbox{boxE2}{
  enhanced,
  colback=white,
  colframe=black,
  boxrule=0.6pt,               
  sharp corners, width=0.43\textwidth,
  height=0.31\textwidth, 
 boxrule = 0pt, % clearing the default rule
    borderline = {0.75pt}{0pt}, % outer line
    borderline = {0.75pt}{0.06cm}, % inner line,
}
\usepackage{musicography}

\makeatletter
\newcommand{\fmarki}{$\sharp$}
\newcommand{\fmarkii}{\ensuremath{\natural}}
\newcommand{\fmarkiii}{\ensuremath{\flat}}
\newcommand{\fmarkiv}{\ensuremath{\mathsection}}
\newcommand{\fmarkv}{\ensuremath{\mathparagraph}}
\newcommand{\fmarkvi}{\ensuremath{\|}}
\newcommand{\fmarkvii}{**}
\newcommand{\fmarkviii}{\ensuremath{\dagger\dagger}}
\newcommand{\fmarkix}{\ensuremath{\ddagger\ddagger}}
                
\def\@fnsymbol#1{{\ifcase#1\or \fmarki\or \fmarkii\or \fmarkiii\or \fmarkiv\or \fmarkv\or \fmarkvi\or \fmarkvii\or \fmarkviii\or \fmarkix \else\@ctrerr\fi}}
\makeatother

\begin{document}

\title{Infinite-dimensional symmetries in plane wave spacetimes}

\author{Emilie Despontin}
\email{emilie.despontin@ulb.be}

\affiliation{Physique Théorique et Mathématique and International Solvay Institutes,\\
Université Libre de Bruxelles (ULB), C.P. 231, 1050 Brussels, Belgium.}
\author{Stéphane Detournay}
\email{sdetourn@ulb.ac.be}
\affiliation{Physique Théorique et Mathématique and International Solvay Institutes,\\
Université Libre de Bruxelles (ULB), C.P. 231, 1050 Brussels, Belgium.}

\author{Dima Fontaine}
\email{dima.fontaine@ulb.be}
\affiliation{Physique Théorique et Mathématique and International Solvay Institutes,\\
Université Libre de Bruxelles (ULB), C.P. 231, 1050 Brussels, Belgium.}

\begin{abstract}
We study the asymptotic symmetries of the Nappi-Witten spacetime in four dimensions, a plane wave arising as the Penrose limit of AdS$_2\times S^2$. Imposing suitable boundary conditions at large transverse distance, we uncover a new infinite-dimensional symmetry algebra allowing for non-trivial central extensions. 
The corresponding phase space encompasses the most general four-dimensional pp-wave metric, including in particular the Penrose limit of Kerr black holes.
\end{abstract}

\maketitle

\setlength\parindent{0pt}

\textbf{Introduction --}
\textit{Plane-fronted waves with parallel propagation}, or \textit{pp-waves}, form a class of geometries defined by the existence of covariantly constant null vector (see \textit{e.g.} \cite{blau2011plane} for a pedagogical introduction to pp-waves). Plane waves are a subclass of pp-waves defined, in Brinkmann coordinates, by
\begin{equation}
    ds^2= 2 du dv + A_{ij}(u) x^i x^j du^2 + (dx^i)^2. \label{Eq:PlaneWaveIntroMetric}
\end{equation}
They are the generic result of a Penrose limit \cite{Marolf_2002, Hubeny_2002}: starting from a null geodesic $\gamma$ in an arbitrary spacetime $\mathcal{M}$, the Penrose limit of $(\mathcal{M}, \gamma)$ is the spacetime $\mathcal{M}_\gamma$ consisting of the infinitesimal neighbourhood of $\gamma$, now blown-up to infinity. This procedure can be generalised to include supergravity fields \cite{Guven_2000}. Plane waves have been put under the spotlight more than two decades ago. These backgrounds yield solvable string theories \cite{Horowitz1990,Horowitz_1994} and in particular, the maximally-supersymmetric plane wave background of type IIB superstring theory was shown to arise as the Penrose(-Güven) limit of AdS$_5 \times S^5$ \cite{Metsaev_2002,metsaev2002exactly}. This led to the geometrical limit of the AdS/CFT correspondence known as the BMN limit \cite{Berenstein:2002jq, Sadri_2004}, relating string theory on this plane wave background to a subsector of $\mathcal{N}=4$ SYM theory. The causal boundary of this plane wave $-$ and actually of all plane waves with non-positive-definite wave profiles, in any dimension $-$ is a one-dimensional null line parametrised by $u$ \cite{Marolf_2002,Hubeny_2002}.\\

Holography in AdS spacetimes involves a bulk theory of gravity and a conformal field theory on its conformal boundary \cite{Maldacena_1999,Gubser_1998,witten1998anti,Aharony_2000}. However, there are several examples where infinite-dimensional symmetries arise on different surfaces, such as in the near-horizon region of black holes \cite{Strominger_1998,Carlip_1999,Cveti__1998,Carlip_2002,Guica_2009}. More recently, it was shown that the eikonal regime of the quasinormal mode spectrum of black holes falls into highest-weight representations of a conformal SL$(2, \mathbb{R})$ symmetry algebra appearing in their near-photon ring region \cite{Gralla:2020srx,Hadar:2022xag}. As the photon ring is essentially the locus of nearly-bound unstable null geodesics, it was also shown that the eikonal regime quasinormal mode spectrum could be retrieved by solving the wave equation on the plane wave background resulting from the Penrose limit at the photon ring \cite{fransen2023quasinormal}. This was also recently generalised in \cite{kapec2024ppwaveshiddensymmetriesblack} to the highly-damped quasinormal mode spectrum of the Schwarzschild black hole by zooming onto the horizon.\\

In this letter, we investigate the asymptotic symmetries of plane wave spacetimes. 
% \SD What has been done before, state of the art -> see citations to hep-th/0306131
Our focus will be the four-dimensional version of the BMN background, known as the \textit{Nappi-Witten spacetime} \cite{Nappi_1993}, defined by the wave profile $A_{ij}= -\delta_{ij}/4,$ $i,j=1,2$. This background is described by a Wess-Zumino-Witten (WZW) model built on the centrally-extended two-dimensional Euclidean algebra, and can also be obtained as the Penrose limit of AdS$_2\times S^2$, which is itself the near-horizon region of near-extremal Reissner-Nordström black holes. Here, the \textit{asymptotic region} of interest is not the causal boundary of the plane wave, but rather the limit $r=(x^i x_i)^{1/2}\to \infty$, which drives the transverse coordinates of the plane wave \eqref{Eq:PlaneWaveIntroMetric} to infinity. This corresponds to the $x_1, x_2 \to +\infty$ limit associated with quasinormal mode boundary conditions posed in \cite{fransen2023quasinormal, kapec2024ppwaveshiddensymmetriesblack}. In this regime, the asymptotic symmetries capture a potential infinite-dimensional symmetry enhancement in the near-ring region of AdS$_2 \times S^2$, or equivalently, the neighbourhood of null geodesics in the near-horizon region of extremal Reissner-Nordström black holes. We will present boundary conditions encompassing the Nappi-Witten spacetime, finite charge excitations thereof, and more general plane waves \eqref{Eq:PlaneWaveIntroMetric}. We will then identify their asymptotic symmetries, which form an infinite-dimensional algebra that, to our knowledge, has not appeared elsewhere.\\

\textbf{The Nappi-Witten model --}
The Nappi-Witten model is based on the metric \eqref{Eq:PlaneWaveIntroMetric} for which we introduce the change of coordinates
\begin{equation}
    x_1 = r \cos \theta, \qquad x_2= r \sin \theta ,
\end{equation}
resulting in
\begin{equation}
    ds^2 = 2 du dv -\dfrac{r^2}{4} du^2 + dr^2 + r^2 d\theta^2   ,\label{Eq:Metricuvrtheta}
\end{equation}
along with a non-trivial Neveu-Schwarz (NS) flux ${H_{ur\theta}=r}$ and a constant dilaton $\phi$. As a WZW model, this model is an  extremum of the NS-NS four-dimensional low-energy effective action
\begin{equation}
    \mathcal{S}= \int d^4x \, \sqrt{-g} \, e^{-2\phi}\left(R - \dfrac{1}{12} H^2 + 4 \nabla_\mu \phi \nabla^\mu \phi \right).
    \label{Eq:sigmamodel}
\end{equation}
The associated equations of motion $-$ or equivalently, in the WZW language, satisfying the one-loop beta function equations $-$ yield
\begin{equation}
\begin{aligned}
    &R - \dfrac{1}{12} H^2 -4 \nabla_\mu \phi \nabla^\mu \phi + 4 \nabla_\mu \nabla^\mu\phi =0, \\
    &R_{\mu \nu}= \dfrac14 H_{\mu \nu}^2 - 2 \nabla_\mu \nabla_\nu \phi,\\
    &\nabla_\mu H^{\mu \nu \rho} =2 \nabla_\mu \phi H^{\mu \nu \rho}.
\end{aligned}
\end{equation}

\textit{Isometries --} 
The four-dimensional non-semisimple algebra $\mathfrak{nw}_4$ upon which this group manifold is constructed is defined by the generators $\{P_1,P_2,J,T\}$, whose commutation relations are given by
\begin{equation}
    \label{Eq:NW4_algebra}
   \hspace{-9 pt} [J,P_i] = \varepsilon_{ij}P_j , \!\! \quad [P_i,P_j] = \varepsilon_{ij}T , \! \! \quad [T,P_i]\!=\![T,J] = 0.
\end{equation}
Notice that it is a central extension of the two-dimensional Euclidean algebra, with $T$ acting as the central element, $P_i$ the temporal and spatial translation generators and $J$ the one of rotations. The isometries of \eqref{Eq:Metricuvrtheta} are
\begin{align}
    \label{KVNW}
	\xi_1 &=  -\dfrac r 2 \sin(\theta - \dfrac u 2) \partial_v + \cos(\theta - \dfrac u 2)\partial_r - \dfrac1r \sin(\theta - \dfrac u 2)\partial_\theta,\vspace{0.2cm}\nonumber\\
	\xi_2 &= \dfrac r 2 \cos(\theta - \dfrac u 2) \partial_v + \sin(\theta - \dfrac u 2)\partial_r + \dfrac1r \cos(\theta - \dfrac u 2)\partial_\theta,\vspace{0.2cm}\nonumber\\
	\xi_3 &= -\dfrac r 2 \cos(\theta + \dfrac u 2) \partial_v + \sin(\theta + \dfrac u 2)\partial_r + \dfrac1r \cos(\theta + \dfrac u 2)\partial_\theta,\vspace{0.2cm}\nonumber\\
	\xi_4 &= \dfrac r 2 \sin(\theta + \dfrac u 2) \partial_v + \cos(\theta + \dfrac u 2)\partial_r - \dfrac1r \sin(\theta + \dfrac u 2)\partial_\theta,\vspace{0.2cm}\nonumber\\
	\xi_5 &= \partial_\theta,\hspace{1.5cm}\quad\xi_6 = \partial_v,\hspace{1.5cm} \quad
	\xi_7 = \partial_u.
    \end{align}
Clearly, $\partial_v$ is the center of the seven-dimensional algebra generated by these vector fields. This algebra can be seen as a double copy (left and right-moving) of $\mathfrak{nw}_4$, with the centers of each copy identified
\begin{equation}
\begin{alignedat}{2}
    &P_1^+ = \xi_2,            &\hspace{1cm} &P_1^- = \xi_4, \\
    &P_2^+ = \xi_1,            &\hspace{1cm}  &P_2^- = \xi_3, \\
    &T     = -\xi_6,           &\hspace{1cm}  &T     = -\xi_6, \\
    &J_{\rm N.W.}^+   =  \dfrac{1}{2}\xi_5 - \xi_7, &\hspace{1cm}  &J_{\rm N.W.}^-   =-  \dfrac{1}{2}\xi_5 - \xi_7.
\end{alignedat}
\end{equation}

\textit{Penrose limit of AdS$_2\times S^2$ --}
The spacetime described by the metric \eqref{Eq:Metricuvrtheta} is actually the Penrose limit \cite{penrose1976any} of ${\text{AdS}_2 \times S^2}$, see \textit{e.g.} \cite{Berenstein:2002jq,Blau_2002}. When the radii of AdS$_2$ and $S^2$ are equal, the metric of AdS$_2 \times S^2$ is given by
\begin{equation}
    ds^2=R^2 \left(-dt^2 \cosh^2\rho+d\rho^2 + d\phi^2 \sin^2\tilde{\theta} + d\tilde{\theta}^2\right).
\end{equation}
The Penrose limit consists on zooming into a geodesic sitting at $\rho=\tilde{\theta}=0$, moving along the $\phi-$direction, which is achieved by considering the coordinates
\begin{equation}
    t=\frac u2-\frac{v}{R^2}, \: \phi = \frac u2 + \frac{v}{R^2}, \: \rho = \frac{r \cos\theta}{R}, \: \tilde{\theta} = \frac{r \sin \theta}{R},
\end{equation}
and taking the limit $R\to\infty$. We end up with the metric \eqref{Eq:Metricuvrtheta} and an enhanced isometry group after the İnönü-Wigner contraction of $SO(1,2) \times SO(3)$ \cite{Blau_2002}.\\

\textit{Relation to Carroll symmetries --}
 It was pointed out in \cite{Duval_2017} that the isometry group of a generic four-dimensional plane wave is identified as the Carroll group in one less dimension with broken rotations.
In four-dimensional plane waves, the dimension of the isometry group is always equal to or larger than five. If the plane wave is conformally flat, then this number is brought to six and the Carroll rotations are restored. Depending on the wave profile, this number can be brought to seven. In the language of \cite{afshar2024classificationconformalcarrollalgebras}, we can show that the isometry algebra of the Nappi-Witten spacetime is isomorphic to the seven-dimensional $d=3, \,z=0$ type-D minimal extension of the Carroll algebra. What is more, on a fixed $u= k \pi/2$ surface (which are null hypersurfaces orthogonal to the causal boundary of the plane wave), one can check that the seven isometries are projected on the $d=3$ type-K Carroll algebra. The infinite-dimensional algebra that we present in this letter is however \textit{not} isomorphic to any infinite-dimensional conformal extension of the Carroll algebra. The bulk realisation of conformal Carroll algebras in plane waves is discussed in \cite{Despontin:2025dog}.\\

\textbf{Boundary conditions and symmetry algebra --}
To generate boundary conditions (BCs) associated to metric \eqref{Eq:Metricuvrtheta}, we introduce a conical defect (CD) in the geometry by imposing periodicity for the $u$ (acting as a time coordinate) and $\theta$ coordinates. Conical defects provide a solution of physical interest (as they model point-particle energy distributions \cite{deser1984three}) from which we can extract fall-offs for the metric, as was for example done in the seminal work \cite{brown1986central}. We impose
\begin{equation}
        u\sim u-2\pi A,\qquad
        \theta\sim \theta + 2\pi \alpha,
\end{equation}
with $A$ and $\alpha$ two constants. By defining new coordinates $u$ and $\theta$ that are respectively not periodic and $2\pi-$periodic, we find the following class of geometries\footnote{This metric was derived with Gim Seng Ng.}:
\begin{equation}
\begin{aligned}
    ds^{\hspace{1pt}2}_{\rm C.D.}  =& \frac{1}{\alpha^2-A^2} dr^2 + r^2 d\theta^2 - \frac14 r^2du^2 \\
    & \hspace{80 pt}+ 2\alpha dudv -4A d\theta dv. \label{Eq:metricNappiWittenCD}
    \end{aligned}
\end{equation}
This family of metrics is governed by both $A$ and $\alpha$ parameters, for which we recover the Nappi-Witten metric when $A=0$ and $\alpha=1$. When $A=0$, we can easily see that the geometry of the $(r,\theta)-$plane is that of a cone. By defining $r\to \alpha r$, we have
\begin{equation}
    ds^{\hspace{1pt}2}_{r,\theta} = dr^2 + \alpha^2 r^2 d\theta^2, 
\end{equation}
which corresponds to a cone whose conical defect is governed by $\alpha$. The $r-$coordinate measures the distance from the geometric singularity.  %Moreover, we realise that this family of metrics possesses a discrete symmetry $(\alpha, A, v)\to (-\alpha,-A,-v)$. 
Performing the same coordinate transformations on the gauge field, one obtains
\begin{equation}
    \mathbf{B}^{\rm C.D.} = \frac{\alpha u - 2 A \theta}{\alpha + A} r \left(dr\wedge d\theta - \frac 12 du \wedge dr\right).
    \label{Eq:BCD}
\end{equation}

Acting on the CD with the Lie derivative along the exact Killing vectors \eqref{KVNW}, one generates BCs, obtained after expanding this variation in the asymptotic region defined by taking $r\to \infty$. This yields
\vspace{-0.2cm}
\begin{align}
    g_{uu}        &= -\frac{r^2}{4} + \mathcal{O}(r),      \quad g_{uv}        = \mathcal{O}(1),       \quad g_{vv}      = 0, \nonumber\\
    g_{\theta\theta} &= r^2 + \mathcal{O}(r) , \qquad\, g_{rr} = \mathcal{O}(1),                     \quad g_{vr}        = \mathcal{O}(1/r^2),    \nonumber\\
    g_{ur}        &= \mathcal{O}(1)
    ,                     \qquad\qquad\,\, g_{v\theta}   = \mathcal{O}(1),        \,\,\,\,\,g_{u\theta}   = \mathcal{O}(r),
    \label{Eq:BCNappiWitten}
\end{align}
where the $u$ and $\theta$ dependencies of the arbitrary functions have been kept implicit, while they are independent in $v$. We also generate BCs for the $\mathbf{B}-$field by acting on the conical defect gauge field $\mathbf{B}^{\rm C.D.}$ with the generalised Killing couples $(\xi_i, \Lambda_i)$\footnote{They can be found in the appended Mathematica file.}:
\begin{equation}
     B_{\mu \nu}^{'\text{C.D.}} = \left(\mathcal{L}_{\xi} \mathbf{B}^{\text{C.D.}} \right)_{\mu \nu} + (d\mathbf{\Lambda})_{\mu \nu},
 \end{equation}
 from which we find the following boundary conditions:
\begin{equation}
\begin{aligned}
    \mathbf{B} &= \left(ur + \mathcal{O}(r)\right)du\wedge dr + \left(\frac12 ur + \mathcal{O}(r)\right) dr\wedge d\theta \\& \hspace{2.5 cm}  + \mathcal{O}(r) du\wedge d\theta.
\end{aligned}
\end{equation}

\textit{Asymptotic Killing vectors --}
Now that we have a set of BCs, we look for the isometries preserving this asymptotic behaviour. We focus on the variation of those BCs under the effect of an arbitrary diffeomorphism $\xi$, whose components are expanded in power series of the radial coordinate $r$:
\begin{equation}
    \xi^\mu = \sum_{n=-\infty}^{+ \infty} r^n \xi^\mu_n(u,v,\theta).
\end{equation}
We manage to constrain orders of the power series by asking that each boundary condition is preserved, \textit{i.e.} 
\begin{equation}
    (\mathcal{L}_\xi g)_{\mu \nu} = \mathcal{O}\left(r^{p_{\mu \nu}}\right),
\end{equation}
with $g_{\mu \nu}=g_{\mu \nu}^\text{NW} + \mathcal{O}(r^{p_{\mu \nu}})$, $g_{\mu \nu}^\text{NW}$ being the background metric. We obtain
\begin{equation}
    \begin{cases}
        \xi^u = A + \mathcal{O}(1/r), \\
        \xi^v = g_1 (u,\theta) r + g_2(u,\theta)+\mathcal{O}(1/r), \\
        \xi^r = h_1 (u,\theta) + \mathcal{O}(1/r),\\
        \xi^\theta = B + m_1(u,\theta)/r + \mathcal{O}(1/r^2).
    \end{cases}    \label{Eq:KillingAsymptotiqueNW}
\end{equation}
Although the allowed powers of $r$ and coordinate dependences are quite constrained, there is no emergent relation between the different coefficients of the power series that would allow us to specify the behaviour of the arbitrary functions appearing in our diffeomorphism.
Therefore, we will use our knowledge of the exact Killing vectors to determine relations between some of the arbitrary functions. When writing the exact Killing vectors of the Nappi-Witten spacetime in the form of \eqref{Eq:KillingAsymptotiqueNW}, one can see that
\begin{equation}
    \begin{cases}
        4 \partial_u g_1(u,\theta) = - \partial_\theta m_1(u,\theta), \\
        \partial_\theta g_1 (u,\theta) = - \partial_u m_1 (u,\theta).
    \end{cases}
    \label{Eq:ObservationsKilling}
\end{equation}
We will assume that these relations also hold for our general diffeomorphism in the asymptotic region. From these relations, we obtain wave equations that encourage us to work in \textit{light-cone coordinates}, defined by
\begin{equation}
    x^\pm = \theta \pm \frac u 2.
\end{equation}
The wave equations obtained through \eqref{Eq:ObservationsKilling} allow us to expand the arbitrary functions $g_1(u,\theta)$ and $m_1(u,\theta)$ into left- and right-moving modes and using Eq. \eqref{Eq:ObservationsKilling}, we end up with the asymptotic diffeomorphism
\begin{align}
\label{Eq:diffeo}
    \xi =& \left[\frac A 2 + B + \frac 2 r \left(-g_+(x^+)+g_-(x^-)\right)\right]\partial_+ \nonumber\\
    &+ \left[-\frac A 2 + B + \frac 2 r \left(-g_+(x^+)+g_-(x^-)\right)\right]\partial_- \nonumber\\
    &+ \left[ r\left(g_+(x^+)+g_-(x^-) \right) + g_2(x^+,x^-)\right]\partial_v \nonumber\\
    &+ \left[ 2\left(\partial_+ g_+(x^+)- \partial_-g_-(x^-) \right)\right]\partial_r.
\end{align}
We now Fourier expand these generators, with the coordinates $x^\pm$ inheriting the $\theta-$periodicity. We extract three infinite towers of generators and two discrete ones:
\begin{align} 
\label{Eq:AKV_NW}
        \xi^+_n &= r e^{i n x^+} \partial_v + 2 i n e^{i n x^+} \partial_r - \dfrac2r e^{i n x^+}\partial_+ - \dfrac2r e^{i n x^+}\partial_-\vspace{0.2cm}\nonumber \\
        \xi^-_n &= r e^{i n x^-} \partial_v - 2 i n e^{i n x^-} \partial_r + \dfrac2r e^{i n x^-}\partial_+ + \dfrac2r e^{i n x^-}\partial_-, \nonumber\\
        J^+ &= \partial_+, \quad
        J^- = \partial_-, \quad P_{n,m} = e^{i n x^+}e^{i m x^-} \partial_v.
    \end{align}
The exact Nappi-Witten symmetries can be expressed in terms of these generators:
\begin{equation}
    \begin{aligned}
        &P_1^+= \frac14 (\xi_{-1}^- +\xi_1^-),            &\hspace{0.2cm} &P_1^- = \frac{-i}{4}(\xi_1^+ - \xi_{-1}^+), \\
    &P_2^+ = \frac{-i}{4}(\xi_{-1}^-  - \xi_1^-),            &\hspace{0.2cm}  &P_2^- = -\frac14 (\xi_1^+ + \xi_{-1}^+), \\
    &T     = -P_{0,0},           &\hspace{0.2cm}  &T= -P_{0,0}, \\
    &J_{\rm N.W.}^+   =  J^-, &\hspace{0.2cm}  &J_{\rm N.W.}^-   =- J^+.
    \end{aligned}
\end{equation}
Moreover, the asymptotic generators form a closed algebra with the following non-zero commutation relations

\begin{center}
\begin{boxE}
\vspace{-0.4cm}
\begin{align}
    &  \hspace{-0.4cm} [\xi^+_m, \xi^+_n] =  4 i (m-n) P_{m+n,0},\nonumber\\
    &  \hspace{-0.4cm} [\xi^-_m, \xi^-_n] =  -4 i (m-n) P_{0,m+n},\nonumber\\
    &   \hspace{-0.4cm} [J^\pm,\xi^\pm_n] = i n \xi^\pm_{n},\nonumber\\
    &  \hspace{-0.4cm}  [J^+, P_{m,n}] = i m P_{m,n},\nonumber\\
    & \hspace{-0.4cm}   [J^-, P_{m,n}] = i n P_{m,n},\label{Eq:AsymptoticAlgebra}
\end{align}
\end{boxE}
\end{center}
in the asymptotic $r\to \infty$ limit. One can verify that the commutation relations (\ref{Eq:AsymptoticAlgebra}) satisfy the Jacobi identity and therefore define a Lie algebra.
It admits the following central extensions

\begin{center}
\begin{boxE2}
\vspace{-0.4cm}
\begin{align}
    & \hspace{-0.34cm}  [\xi^+_m, \xi^+_n] =  4 i (m-n) P_{m+n,0}+ k(n)\delta_{n+m,0},\nonumber\\
    & \hspace{-0.34cm}  [\xi^-_m, \xi^-_n] = - 4 i (m-n) P_{0,m+n}+ \tilde{k}(n)\delta_{n+m,0},\nonumber\\
    & \hspace{-0.34cm}  [J^\pm,\xi^\pm_n] = i n \xi^\pm_{n},\nonumber\\
    &  \hspace{-0.34cm} [J^+, P_{m,n}] = i m P_{m,n},\nonumber\\
    & \hspace{-0.34cm}  [J^-, P_{m,n}] = i n P_{m,n},\nonumber\\
    & \hspace{-0.34cm}  [\xi^+_n, P_{m,0}] = (c_3 n^3 +c_1 n)\delta_{n+m,0},\nonumber\\
    & \hspace{-0.34cm}  [\xi^-_n, P_{0,m}] = (\tilde{c}_3 n^3 +\tilde{c}_1 n)\delta_{n+m,0},\nonumber\\
    & \hspace{-0.34cm}  [\xi^+_p, P_{m,n}] = h(m,n)\delta_{p+m,0},\hspace{0.2cm}n\neq 0,\nonumber\\
     & \hspace{-0.34cm}  [\xi^-_p, P_{n,m}] = \tilde{h}(n,m)\delta_{p+m,0},\hspace{0.2cm} n\neq 0,
\end{align}
\end{boxE2}
\end{center}
with $c_1$, $c_3$, $\tilde{c}_1$and $\tilde{c}_3$ constant and $k(n)$, $\tilde{k}(n)$, $h(n,m)$ and $\tilde{h}(m,n)$ some odd functions in $n$, which cannot be absorbed in a redefinition of the generators. We can establish the correspondence with the Carroll algebra by introducing
\begin{equation}
    A_n = i \left(\xi^-_{-n}+ \xi^+_n\right) \quad \text{and} \quad B_n = \left(\xi^-_{-n}- \xi^+_{n}\right),
\end{equation}
such that we recover the three-dimensional Carroll generators taking
\begin{equation}
\begin{aligned}
    C_1    &\equiv \dfrac{i}{2} \xi_2     = -\frac{1}{2} A_1,       &\quad 
    P_1    &\equiv \dfrac{i}{2} \xi_3     = \frac{1}{2} A_{-1}, \\
    C_2    &\equiv -\dfrac{i}{2} \xi_1    = \frac{1}{2} B_1,        &\quad 
    P_2    &\equiv \dfrac{i}{2} \xi_4     = -\frac{1}{2} B_{-1}, \\
    J_{12}  &\equiv \xi_5                  = J^+ + J^-,             &\quad 
     H      &\equiv \xi_6                  = P_{00},\\
    D     &\equiv \xi_7                  = \frac12 (J^+ - J^-), 
\end{aligned}
\end{equation}
with $C_1, C_2$ the two Carroll boosts, $P_1,P_2$ the two spatial translations, $H$ the time translation, $J_{12}$ the rotation, as well as an additional dilation $D$. We also exhibited the correspondence with the global isometries \eqref{KVNW}.\\

\textit{Generalisation of boundary conditions --}
Furthermore, we show that our BCs include the most general pp-wave metric. In Brinkmann coordinates, this family of metrics takes the following form \cite{blau2011lecture}
\begin{equation}
\begin{aligned}
    ds^2=&\, 2 dudv+ K(u,x^b)du^2+ dx_1^2 + dx_2^2\\
    & \hspace{85 pt}+2A_a(u,x^b)dx^a du,
\end{aligned}
\end{equation}
where the indices $a,b=1,2$. Additionally, plane waves are a particular type of pp-wave for which 
\begin{equation}
    ds^2=2 du dv + A_{ab}(u) x^a x^b du^2 + dx_1^2+dx_2^2.
\end{equation}
Letting $x_1=r\cos\theta$ and $x_2=r\sin\theta$, one has
\begin{equation}
\begin{aligned}
    ds^2&=2dudv + K(u,r,\theta) du^2+ dr^2+ r^2d\theta^2\\
    & \hspace{-5pt}+ \left[2A_1(u,r,\theta) \cos\theta+2 A_2 (u,r,\theta) \sin\theta\right] dudr\\
    & \hspace{-5pt}+ \left[-A_1(u,r,\theta)  r\sin\theta +A_2(u,r,\theta)  r \cos\theta\right] d\theta du .
\end{aligned}
\end{equation}
This is contained in our BCs, as long as $K(u,r,\theta)$ is at most quadratic in $r$, while allowing $g_{uu} = \mathcal{O}(r^2)$, instead of $g_{uu}=-r^2/4+\mathcal{O}(r)$. One can also verify that our asymptotic diffeomorphism \eqref{Eq:diffeo} preserves these more relaxed BCs, giving rise to the same behaviour of the charges\footnote{See appended mathematica file for the details of the computations} as computed in the following section. Let us mention that this is an interesting result as one can already find in the literature some known examples satisfying these BCs, see \textit{e.g.} the Penrose limit of the Schwarzschild metric \cite{blau2011lecture}, the Penrose limit of the Kerr black hole, as well as the Penrose limit of the near-horizon near-extremal limit (near-NHEK) \cite{fransen2023quasinormal}. For instance, the Penrose limit of the Schwarzschild spacetime is given by the (Einstein) metric
\begin{equation}
    ds^2 = 2 du dv + \dfrac{1}{3 M^2} (x^2-y^2) du^2 + dx^2 + dy^2.
\end{equation}
Going to polar coordinates, we find 
\begin{equation}
    K(u,r,\theta) = \dfrac{r^2 \cos 2 \theta}{3M^2}, \quad A_i(u,r,\theta)=0.
\end{equation}

\textbf{Surface Charges --}
As a next step, we compute the surface charges associated to the (asymptotic) isometries, using the machinery developed in \cite{Iyer_1994, BarnichBrandt} (see \cite{Compere:2007vx,Detournay_2013} and the appended Mathematica file for explicit expressions, along with the SurfaceCharges package\footnote{\url{https://ptm.ulb.be/gcompere/package.html}}). We first focus on the surface charges of the global isometries after introducing the conical defect. Then we turn to the ones of the asymptotic isometries. We will explore two scenarios, both for which those charges are finite.\\

\textit{Charges associated to the conical defect --}
 Out of the seven global isometries \eqref{KVNW}, three are preserved after the introduction of the conical defect: $\xi_5,\xi_6$ and $\xi_7$. Their associated surface charges are given by
\begin{equation}
    \begin{cases}
        \delta Q_5 = \frac{1}{4\pi G} (\alpha \delta A + A \delta \alpha), \\
        \delta Q_6 = 0, \\
        \delta Q_7 = \frac{-1}{8\pi G} (\alpha\delta \alpha + A \delta A),
    \end{cases}
    \label{chargesconical}
\end{equation}
which are finite and integrable. \\

\textit{An abelian charge algebra --}
First, we consider the case where all BCs do not depend on $v$, at \textit{any order}. We introduce a parameter $p$ measuring the deviation in phase space with respect to our reference solution. The $p-$form contribution to the infinitesimal charges is identical for all the asymptotic generalised Killing symmetry couples and can be shown to vanish on-shell. The only contribution to the infinitesimal charges is gravitational. We set a phase space condition
\begin{equation}
    g_{rr}^{(0)} g_{v+}^{(0)} g_{v-}^{(0)} = -1,
    \label{Eq:Condgrrgvpgvm}
\end{equation}
where we use the notation
\begin{equation}
    g_{\mu \nu} = r^n g_{\mu \nu}^{(0)} + r^{n-1} g_{\mu \nu}^{(-1)}+\dots
\end{equation}
with $n$ the highest power in $r$ in a given metric component, and the metric corresponds to \eqref{Eq:BCNappiWitten} expressed in the light-cone coordinates. The infinitesimal charges are integrable and given by
\begin{align}
    Q_{J^+} &= \frac{-1}{16\pi G} \int dv \int_0^{2\pi} d\theta \, \sqrt{\frac{-{g_{v+}^{(0)}}^3}{g_{rr}^{(0)}g_{v-}^{(0)}}}, \\
    Q_{J^-} &=\frac{1}{16\pi G} \int dv \int_0^{2\pi} d\theta \, \sqrt{\frac{-{g_{v-}^{(0)}}^3}{g_{rr}^{(0)}g_{v+}^{(0)}}}.
\end{align}
Moreover, the EOMs yield
\begin{equation}
    \partial_\pm g_{v-}^{(0)} = 0\quad \text{and} \quad \partial_\pm g_{v+}^{(0)} = 0,
\end{equation}
resulting in an abelian algebra
\begin{equation}
    \{Q_{J^+},Q_{J^-}\}=0.
\end{equation}

\textit{Towards an infinite-dimensional algebra --}
From the beginning, we considered our BCs to be independent of $v$. However, the Killing equations allow us to keep a dependence on this coordinate at subleading orders. These slightly modified BCs still yield
\begin{equation}
    \delta Q_{P_{m,n}}=0
\end{equation}
when $r\to\infty$, but have excited the infinitesimal charges associated to $\xi^\pm_m$.  By putting them on-shell, the relevant equation of motion is \begin{equation}
    \label{Eq:Onshell}
    g_{rr}^{(-1)}=0,
\end{equation}
and
\begin{align}
    \delta Q_{\xi^\pm_m} =& \frac{\pm\delta p}{32\pi G} {g_{rr}^{(0)}}^{2}T'_m\left(x^\pm\right) \Bigg[ \frac{2}{g_{rr}^{(0)}} \partial_p\partial_v\left(g_{v-}^{(-1)} - g_{v+}^{(-1)}\right)  \nonumber \\
    & \hspace{-1cm}+\left(g_{v-}^{(0)}  \partial_v g_{v+}^{(-1)} + g_{v+}^{(0)} \partial_v g_{v-}^{(-1)}\right) \partial_p\left(g_{v+}^{(0)} - g_{v-}^{(0)}\right) \Bigg]. 
    \label{Eq:InfChargeXi+}
\end{align}
We find that this tower of infinitesimal charges is non-integrable. Such non-integrability may signal the presence of a physical flux through the boundary\footnote{While non-integrability can indicate a physical flux, this is not always the case. A more precise characterization could be obtained by computing the Weyl scalars in the Newman–Penrose formalism, which we leave for future work.}, preventing a direct computation of Poisson brackets as in previous analyses. Modified brackets have been proposed in the literature, most notably the Barnich–Troessaert brackets \cite{Barnich_2011, bosma2023radiative}, originally developed for the $\mathfrak{bms}_4$ algebra, which require an arbitrary choice in splitting the charges into integrable and non-integrable parts. In our case, this procedure yields partially non-antisymmetric results, suggesting that the modified Barnich-Troessaert brackets are not well suited to handle this form of non-integrability. Alternative approaches exist, such as the Wald–Zoupas prescription, which selects a preferred presymplectic potential, or a generalization of the Barnich-Troessaert bracket recently proposed to represent the extended corner algebra \cite{freidel2021canonicalbracketopengravitational, Freidel:2021cbc}. We do not discuss these in the present work and leave that analysis to future endeavours.
Finally, we note that while the charges associated with the $J^\pm$ generators are altered by these boundary conditions, they remain finite.\\

\textbf{Conclusion and discussion} -- 
Starting from the four-dimensional Nappi-Witten spacetime, we have constructed boundary conditions preserved by the infinite-dimensional symmetry algebra \eqref{Eq:AsymptoticAlgebra}. The exact nature and relevance of this algebra in physical settings has yet to be fully explored,
but we close this letter with some observations. 
This algebra describes an interplay between two chiral current-like fields with modes $\xi_n^\pm$ and a bi-chiral field with modes $P_{m,n}^\pm$, with the global generators $J^\pm$ acting as momenta or scaling weights. 
These structures resemble current-like algebras, but where the central extension (normally a constant) is replaced by a field/operator.
Since the $\xi^\pm_m$ alone do not close, the $P_{m,n}^\pm$ generically represent fluxes responsible for charge non-conservation (but vanishing for our choice of boundary conditions). The study of gravitational phase spaces with non-vanishing fluxes, non-integrable charges and state-dependent central extensions is a delicate and timely subject that would require further investigations (see \textit{e.g.} \cite{Barnich_2011, Freidel:2021cbc, Ciambelli:2022cornerproposal, Adami:2020null, Zwikel:2021bondiweyl, Zwikel:2023wfg, Ciambelli:2020ftk}). 
On the other hand, even though field theories with symmetries \eqref{Eq:AsymptoticAlgebra} have not been identified to our knowledge, one could attempt at engineering toy models capturing some of their features. A candidate for a simple example is given by the Lagrangian 
\begin{equation}
    \mathcal{L} = \partial_+ \chi \, \partial_- \chi + \chi \, \partial_+ \phi^+ \, \partial_- \phi^-,
    \label{eq:lagrangian}
\end{equation}
where $\chi (x^+,x^-)$ and $\phi^\pm  (x^\pm)$ are scalar fields (chiral for $\phi^\pm$).
The model is invariant under independent chiral shifts of \(\phi^\pm\): $ \delta \phi^+ = \epsilon^+(x^+), \quad \delta \phi^- = \epsilon^-(x^-), \quad \delta \chi = 0$. The associated 
Noether charges are 
\begin{equation}
 Q_\pm[\epsilon^\pm] = \int_0^{2\pi} d\varphi \, \chi \, \partial_\mp \phi^\mp \, \epsilon^\pm(x^\pm),
   \end{equation}
while their Fourier modes can be identified with $\xi^\pm_m$. The system exhibits primary constraints which renders the canonical analysis non-trivial, requiring the use of Dirac brackets. This then suggests a charge algebra of the form $\{\xi_+^m, \xi_+^n\}_D = 4 i (m - n) P_{m+n,0}$ with $P_{m,n} = \int_0^{2\pi} d\varphi \, \chi(x^+, x^-) e^{i (m x^+ + n x^-)}$, and similarly for the ``$-$" sector. The field $\chi$ can be viewed as a flux mediator, coupling the left- and right-moving field terms $\phi^\pm$ through the interaction $\chi\, \partial_+ \phi^+ \partial_- \phi^-$, while its dynamics is governed by
the equation of motion $2 \partial_+ \partial_- \chi = \partial_+ \phi^+\, \partial_- \phi^-,$ where $\partial_+ \phi^+ \partial_- \phi^-$ plays the role of a flux density.
Thus, $\chi$ can be interpreted physically as a potential field sourced by the flux associated to $\phi^\pm$. Additionally, the action is also invariant under global chiral dilations associated with global charges $J^\pm$. A detailed analysis of this model, including its canonical structure, quantization, the appearance of potential central extensions, 
%(due to normal ordering), 
 partition function, is left for future work, along
with the study of general properties of the symmetry algebra such as representation theory, coadjoint orbits, modular properties and geometric actions. Such a simple action is of course not believed to capture holographic properties of pp-wave spacetimes, no more than a massless scalar two dimensional CFT should describe AdS$_3$ gravity or a Carrollian boson being dual to gravity in flat space. However, it opens the way for studying new classes of potentially solvable field theories whose generic structure could be relevant in a gravity context. 
% Could quantize the system, study central extensions, more generally representation theory of the algebra (hw, indiced), coadjoint orbits, partition function
On another hand, it has been shown \cite{Gimon_2003} that black strings can be embedded in pp-wave spacetimes in five and more dimensions. Hence, a question of interest is how these solutions can be included in higher-dimensional generalisations of our boundary conditions and how this would fit within the surface charges presented in \cite{Compere:2007vx}. Finally, we mentioned in the introduction the intimate relation between Penrose limits and eikonal quasinormal modes, whose wavefunctions are generically obtained in terms of parabolic cylinder functions. It might be interesting to investigate whether the latter play, in the algebra \eqref{Eq:AsymptoticAlgebra}, a role comparable to that of hypergeometric functions for the Virasoro algebra.

\begin{acknowledgments}
\vspace{-0.48cm}
SD is much indebted to Gim Seng Ng for early discussions and collaboration that led to this work. We warmly thank Mukund Rangamani, Marc Henneaux, Arjun Bagchi, Céline Zwikel, Daniel Grumiller, Marc Geiller and Sudipta Dutta for helpful discussions. SD is a Senior Research Associate of the Fonds de la Recherche Scientifique F.R.S.-FNRS (Belgium).
He acknowledges support of the Fonds de la Recherche Scientifique F.R.S.-FNRS (Belgium) through the following projects: CDR project C 60/5 - CDR/OL “Horizon holography : black holes and field theories” (2020-2022), PDR/OL C62/5 ``Black hole horizons: away from conformality'' (2022-2025) and CDR n°40028632 (2025-2026). This work is supported by the F.R.S.-FNRS (Belgium) through convention IISN 4.4514.08 and benefited from the support of the Solvay Family. DF benefits from a FRIA fellowship granted by the F.R.S-FNRS. ED is a Research Fellow
of the Fonds de la Recherche Scientifique F.R.S.-FNRS (Belgium). The authors are members of BLU-ULB, the interfaculty research group focusing on space research at ULB.
\end{acknowledgments}

\bibliography{biblio_uniform}
\end{document}